\documentclass[fleqn,usenatbib]{mnras}
\usepackage{graphicx}	% Including figure files
\usepackage{amssymb}	% Extra maths symbols
\usepackage{epstopdf}
\usepackage{color}
\usepackage{amsmath}
\usepackage{mathtools}
\usepackage{hyperref}% \documentclass[manuscript]{aastex}\def\apj{{\rm ApJ}}                    % Astrophys. J.
\usepackage{perpage}
\def\apjs{{\rm ApJS}}                  % Astrophys. J. Supl.
\def\apjs{{\rm ApJ}}                   % Astrophys. J.
\def\apjl{{\rm ApJ Let}}               % Astrophys. J. (Letters)
\def\aj{{\rm AJ}}                      % Astron. J.
\def\mnras{{\rm MNRAS}}                        % Monthly Notices

\def\Msol{\thinspace\hbox{$\hbox{M}_{\odot}$}}
\def\Zsol{\thinspace\hbox{$\hbox{Z}_{\odot}$}}
\def\Dsol{\thinspace\hbox{$\hbox{D}_{\odot}$}}

\def\ie{{\it i.e.} }                    % i.e. italicised
\def\eg{{\it e.g.} }                    % e.g. italicised
\def\a4{\hsize 17.0cm \vsize 25.cm}
\newcommand{\der}[2]  { \frac{{\rm d}#1}{{\rm d}#2} }
\newcommand{\derp}[2] { \frac{\partial #1}{\partial #2} }
\newcommand{\dif}     {{\rm d}}

\title{Dust Grain Growth at High Redshift: Starburst-driven CMB-Dark Supershells}

\author[S. Mart\'{\i}nez-Gonz\'alez et al.]
{Sergio Mart\'{\i}nez-Gonz\'alez \thanks{E-mail: sergiomtz@inaoep.mx}$^{1},$ 
Sergiy Silich $^{2},$
Guillermo Tenorio-Tagle$^{2},$
\\
$^{1}$CONACYT-Instituto Nacional de Astrof\'isica, \'Optica y Electr\'onica, AP 51, 
      72000 Puebla, M\'exico\\
$^{2}$Instituto Nacional de Astrof\'\i sica \'Optica y Electr\'onica, AP 51, 
      72000 Puebla, M\'exico\\
}

\date{Accepted July 27th, 2021.}

\pubyear{2021}

\begin{document}
% \label{firstpage}
% \pagerange{\pageref{firstpage}--\pageref{lastpage}}
\maketitle

\begin{abstract}
We present a novel scenario for the growth of dust grains in galaxies at high-redshift ($z\sim6$). In our model, the mechanical feedback from massive star clusters evolving within high-density pre-enriched media allows to pile-up a large amount of matter into 
massive supershells. If the gas metallicity ($\geq\Zsol$), number density ($\geq 10^6$ cm$^{-3}$) and dust-to-gas mass ratio ($\sim 1/150 \times Z$) within the supershell are sufficiently large, such supershells may become optically thick to the starlight emerging from their host star clusters and even to radiation from the Cosmic Microwave Background (CMB). Based on semi-analytic models, we argue that this mechanism, occurring in the case of massive ($\geq 10^7\Msol$) molecular clouds hosting $\geq 10^6\Msol$ star clusters, allows a large mass of gas and dust to acquire a temperature below that of the CMB, whereupon dust grain growth may occur with ease. In galaxies with total stellar mass $M_{*}$, grain growth within supershells may increase the dust mass by $\sim10^6\Msol (M_{*}/10^{8}\Msol)$.
\end{abstract}

          
\section{Introduction}

Violent star formation within the first $\sim10^9$ years of cosmic time ($z \gtrsim 6$) induced a  
super-solar metal enrichment as observed in quasar host galaxies \citep[\eg][]{Jiangetal2007,Caluraetal2014}. 
Furthermore, the detection of massive amounts of dust in the gravitationally-lensed galaxy A2744\_YD4 ($z\sim8.38$) reveals 
that an early dust enrichment occurs during the epoch of cosmic reionisation \citep{Laporteetal2017}, when 
the universe was $\sim200$ Myr old. It has been argued that galaxies at high-redshifts must experience 
a rapid ($\sim$ a few million years long) transition between relatively dust-free and dust-rich 
\citep{Mattssonetal2015}, and that star formation may have proceeded with a top-heavy initial mass 
function \citep{GallandHorth2011,DwekandCherchneff2011}. Several possible dust producers at high redshift 
have been discussed in the literature. For example, the cold envelopes of the most massive AGB stars have been 
regarded as possible dust producers at high-redshift \citep[\eg][]{Valianteetal2009,LesniewskaandMichalowski2019},
particularly in the case of a top-heavy initial mass function \citep{Chiosietal1998}, while 
\citet[][ and references therein]{TodiniandFerrara2001,Indebetouwetal2014} have claimed that a large fraction 
of the early dust enrichment follows from the very fast and highly efficient dust grain condensation in supernova 
ejecta. 

The relative importance of dust formation in supernova ejecta, dust destruction by supernova-driven 
shocks and dust grain growth (by accretion of gas-phase species) in dense and cold molecular clouds (MCs) 
is still matter of intense debate. For instance, \citet{Caluraetal2008,Caluraetal2014},  
confronted the dust destruction rate obtained by \citet{Mckee1989} for SNe exploding in a three-phase
interstellar medium to the dust production rates derived from galactic chemical evolutionary models \citep[see also][]{Dwek1998,Zhukovskaetal2008,Asanoetal2013}. They asserted that dust grain 
growth is a necessary mechanism to account for the presence of dust in galaxies, and to effectively counteract dust 
destruction in supernova-driven shocks. The processes of grain coagulation and grain growth are tightly related, but 
the former has been shown to be unimportant in calculating the evolution of the total dust mass in galaxies 
\citep{Hirashita2012}. More recently, \eg \citet{GallandHorth2018}, have provided supporting evidence in that the dust content in local and 
high-redshift galaxies is consistent with an efficient dust production by supernovae and dust reformation 
shortly after destruction. 

Notwithstanding, \citet{Ferraraetal2016} has noted that the process of grain growth must overcome several complications under 
the conditions prevailing in galaxies at high redshift. First, at $z \gtrsim 6$, the Cosmic Microwave Background (CMB) sets 
the minimum possible grain temperature as $T_{cmb}\gtrsim 20$ K \citep{daCunhaetal2013}, this translates into warmer grain 
surfaces and reduced accretion rates. Second, after the host molecular clouds are disrupted (typically in $\sim10 $ Myr), 
any material accreted onto the grain surfaces will almost immediately be photo-desorbed as the grains are exposed to the 
radiative feedback. A third complication arises from the Coulomb repulsive forces between positively-charged ions and grains in the presence of a strong UV radiation field. Given the aforementioned problems, their view favours the 
notion that the large quantities of dust observed at high redshift must have been readily condensed in supernova 
ejecta. In this paper we will reexamine these objections in the context of starburst-driven superbubbles.

\begin{figure}
\includegraphics[width=\columnwidth]{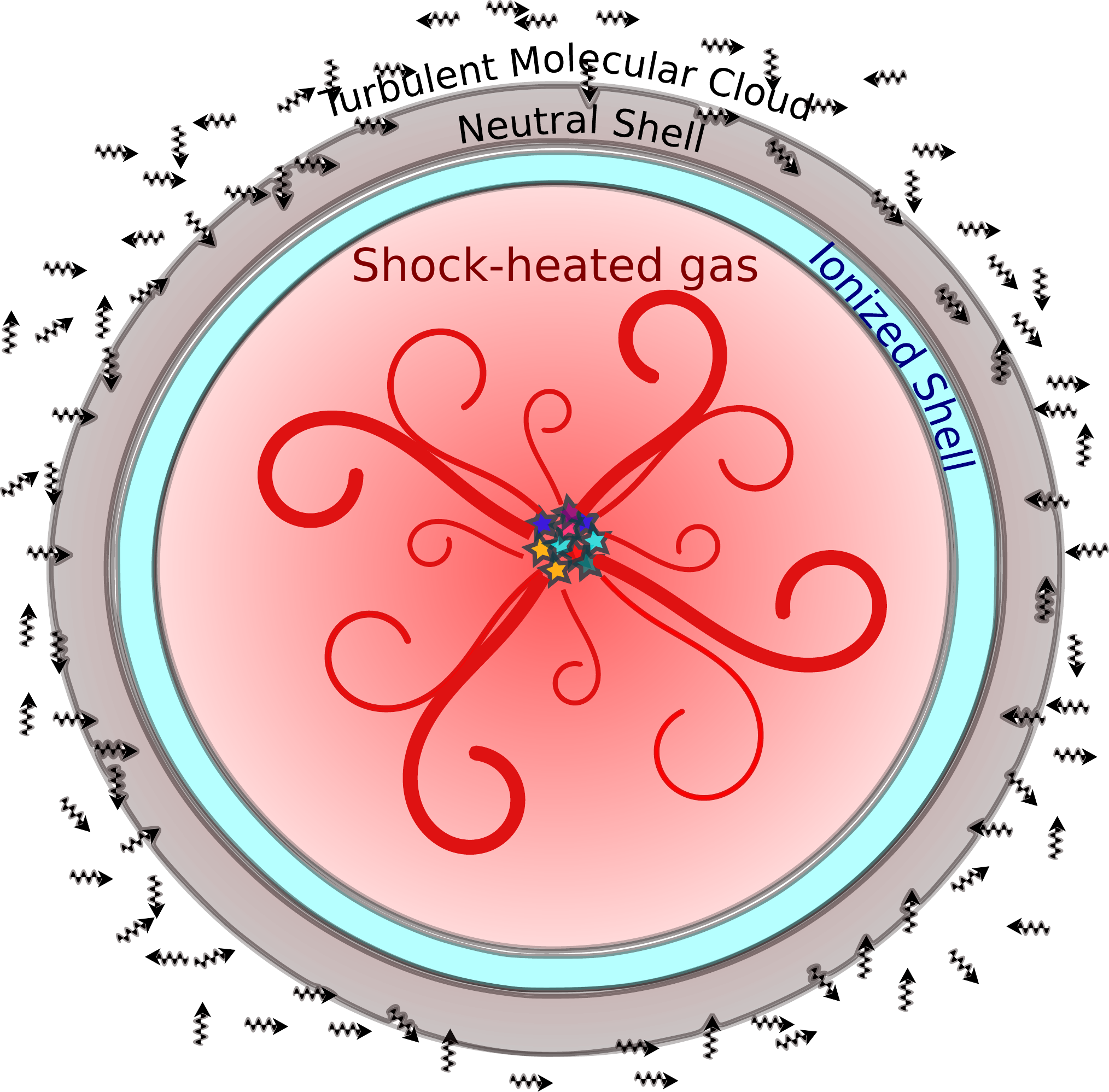}
\caption{Schematic representation of the structure of a starburst-driven superbubble, and the central star cluster (not to scale). Each region of the superbubble is indicated: the region filled with shock-heated gas, the dense/warm ionised supershell, the very dense/cold neutral supershell, and the unshocked turbulent molecular cloud. The bath of CMB photons reaching the neutral supershell is represented by wavy arrows.}
\label{fig:bubble}
\end{figure}

In the ``standard bubble model'', the stellar winds of massive stars, interacting with the surrounding ISM, create a four-distinct-zone 
structure, namely: (i) the central hot and tenuous free wind region, (ii) a (reverse-) shocked wind region separated by a contact discontinuity from (iii) a massive, rapidly-cooling, supershell of swept-up interstellar matter that is able to absorb a large part of the impinging ionising and non-ionising radiation, and (iv) the ambient medium, an MC in our case 
\citep[see][]{Weaveretal1977, MacLowMcCray1988, KooMcKee1992, BKSilich1995}.  

As the gas in the supershell recombines efficiently and dust absorbs large amounts of radiation, it may eventually occur that the ionisation front becomes trapped within the supershell and thus a fifth zone is formed (see Figure \ref{fig:bubble}): (v) an outer dusty neutral layer in the supershell. For example, for a supershell driven by the mechanical feedback of a $10^6 \Msol$ stellar cluster, embedded into a medium with constant density $\sim10^3$ cm$^{-3}$, the ionisation front becomes trapped in $\sim1.5$ Myr \citep{MartinezGonzalezetal2014}, and earlier at higher densities. Following \citet{Weaveretal1977}, we will not distinguish between  regions (i) and (ii), because the radius of the eventually-formed reverse shock is usually much smaller than the radius of the forward shock. Thus, in the following we will refer to them as the shock-heated gas region.

Moreover, radiative cooling induced by dust grains \citep{Whitworth2016} and hydrogen-deuteride (HD) molecules, with the latter efficiently formed in post-shocked gas, may contribute to decrease the neutral supershell temperature to the floor set by the CMB \citep{JohnsonandBromm2006}. 

Furthermore, if the dust column density and the average grain cross section within the neutral supershell are sufficiently large, 
there is an opportunity to trap all incoming CMB radiation before reaching the supershell's ionised part. Consequently, some 
part of the neutral supershell may become self-shielded from the inner starlight and from the outer CMB radiation. Interestingly, 
self-shielding and cooling via adiabatic expansion have been invoked to explain that the outflowing matter in the Boomerang 
Nebula is colder than the current CMB temperature \citep{SahaiandNyman1997,Sahaietal2017}. Our investigation tests the 
feasibility of dust grain growth within the neutral parts of very dense ($\geq 10^6$ cm$^{-3}$), dusty (with dust-to-gas mass ratio $\sim 1/150 \times Z$), and heavily pre-enriched ($\geq\Zsol$), starburst-driven supershells.

The paper is organised as follows: Section \ref{sec:model} describes the superbubble model, including the equations
governing the supershell's ionised and neutral zones, and the main parameters and relevant time-scales. In Section \ref{sec:runs}
we change several parameters to explore the evolution of supershells under different ambient conditions and star cluster masses. 
In Section \ref{sec:growth} we study the feasibility of dust grain growth in supershells at high-redshift. Finally, in Section \ref{sec:conclusions} we summarise our conclusions.

\section{The large-scale evolution}
\label{sec:model} 
\subsection{The Host Molecular Cloud}

We consider a young massive stellar cluster located at the centre of a pre-enriched molecular cloud supported by turbulent 
pressure \citep[][see Appendix \ref{app:turb}]{ElmegreenandEfremov1997,Johnsonetal2015,Caluraetal2015,Elmegreen2017}. We assume 
that the molecular cloud is formed out of material with super-solar metallicity ($Z_{ISM}\geq\Zsol$). The turbulent molecular cloud's assumed gas density distribution is a power-law of the form \citep[\eg][]{Leeetal2015,Raskuttietal2017}

\begin{eqnarray}
\label{eq:powerlaw}
\rho=\rho_{c}\left(\frac{r}{R_{c}}\right)^{-\omega} \,~,\,\,\,\,\, \mbox{for } r\geq R_{c} ~,
\end{eqnarray}

where $r$ is the radial distance, $R_{c}$ is a characteristic length scale or ``core radius'', and 
$\rho_{c}$ is the gas density at $r=R_{c}$. Within the central zone, we assume that the gas density distribution is homogeneous,

\begin{eqnarray}
\label{eq:powerlaw2}
\rho=\rho_{c} \,~,\,\,\,\,\, \mbox{for } r< R_{c} ~.
\end{eqnarray}

The molecular cloud is truncated at a radius $R_{cut}$\(\gg\)$R_{c}$. 

\subsubsection{Superbubble Evolution}

We assume that the kinetic energy injected by massive stars is thermalized in neighboring stellar wind collisions. Once the thermalized gas pressure overcomes the turbulent pressure in the ambient molecular cloud, the shock-heated gas begins to expand. In the spherically-symmetric case, this occurs at $t_0 = 4 \pi R^3_c P_t(R_c) / (3 (\gamma-1) L_{SC}$, where $L_{SC}$ is
the star cluster mechanical luminosity, $P_t(R_c)$ is the molecular gas turbulent pressure at the star cluster edge (see Appendix \ref{app:turb}) and $\gamma = 5/3$ is the ratio of specific heats. The dynamics of the expanding bubble is determined by the set of mass, momentum and energy conservation equations \citep{Weaveretal1977, MacLowMcCray1988, BKSilich1995}. In the case of a homogeneous ambient gas density distribution these equations have a well-known power-law solution. This solution, however, is asymptotic as it does not account for the initial conditions (at $t = 0$ years, the shell radius is zero while the shell velocity is infinite) and also it does not take into consideration the thermalized gas radiative cooling. We add to the Weaver's et al. equations the re-inserted gas radiative cooling and adopt that the ambient gas density distribution is determined by equation \eqref{eq:powerlaw}. The conservation equations that determine the starburst-driven bubble evolution then are
%--------------------------------------------------------------------
\begin{eqnarray}
       \label{eqXa}
      & & \hspace{-1.0cm} 
M_{sh}(r) = \frac{4 \pi \rho_c R^3_c}{3 - \omega}
      \left[\left(\frac{r}{R_c}\right)^{(3-\omega)} - 1\right] ,
      \\[0.2cm] \label{eqXb}
      & & \hspace{-1.0cm} 
      \der{u}{t} = \frac{4 \pi r^2}{M_{sh}(r)} \left[P_{b} - \rho_c u^2
        \left(\frac{R_c}{r}\right)^{\omega} \right] ,
      \\[0.2cm] \label{eqXc}
      & & \hspace{-1.0cm}
\der{E_{b}}{t} = L_{SC} - 4 \pi r^2 P_{b} u - \frac{3 (\dot{M} t)^2
        \Lambda(T_{b},Z_{ISM})}{4 \pi r^3 \eta^2_{ion}} ,
      \\[0.2cm] \label{eqXd}
      & & \hspace{-1.0cm}
\der{r}{t} = u ,     
\end{eqnarray} 
%-------------------------------------------------------------------
where $M_{sh}(r)$ is the supershell's gas mass, the shock-heated gas thermal pressure $P_{b} = 3 (\gamma-1) E_{b} / (4 \pi r^3)$, $\mu_{ion} = 14/11 m_H$ is the mean mass per ion in the mass-loaded, shock-heated plasma with 1 helium atom per each 10 hydrogen atoms, $m_H$ is the mass of the hydrogen atom, $\Lambda(T_{b},Z_{ISM})$ is the \citet{Raymondetal1976} cooling function and 
$\dot{M} = 2 L_{SC} / V_{\infty}^2$ is the mass input rate within the star cluster volume that may include the re-inserted gas mass loading by the residual gas. 

The inside bubble gas mass density and temperature are
%--------------------------------------------------------------------
\begin{eqnarray}
       \label{eqXe}
      & & \hspace{-1.0cm} 
\rho_{b} = \frac{3 \dot{M} t}{4 \pi r^3} ,
      \\[0.2cm] \label{eqXf}
      & & \hspace{-1.0cm} 
T_{b} = \frac{\mu_{a} P_{b}}{k_B \rho_{b}} ,
\end{eqnarray} 
%-------------------------------------------------------------------
where $\mu_{a} = 14/23 m_{H}$ is the mean mass per particle in the completely ionised gas with 1 helium atom per each 10 hydrogen atoms and $k_B$ is the Boltzmann constant. We solve equations (\ref{eqXa}) - (\ref{eqXd}) numerically upon the
assumptions that, at $t = t_0$, the shell radius is slightly larger than the star cluster radius, $r = \alpha R_c$, the shell 
velocity $u = 0$ km s$^{-1}$ and $\alpha = 1.05$. 

In the case of a homogeneous ambient gas density distribution and negligible thermalized gas cooling, the numeric solution rapidly approaches the classic power-law solution.

If the residual gas mass is evenly incorporated into the outflow during a time $t_{ml}$, then the mass-loading 
factor is

\begin{eqnarray}
\label{eq:etaml}
 \eta_{ml} = \frac{M_{MC}(r<R_{c})}{\dot{M}_{w}t_{ml}} ~,
\end{eqnarray}

where $M_{MC}(r<R_{c})$ is

\begin{eqnarray}
\label{eq:Mr0}
 M_{MC}(r<R_{c}) = \frac{4}{3}\pi\rho_{c}R_{c}^3  ~.
\end{eqnarray}

As the outflow is mass-loaded, the adiabatic terminal speed is changed to

\begin{eqnarray}
\label{eq:v_eta}
 V_{\eta,\infty}=V_{\infty}(1+\eta_{ml})^{-1/2}  ~.
\end{eqnarray}

\subsection{The Micro-Physics and the Shell Inner Structure}

The supershell's inner structure is calculated by integrating the following set of 
equations, starting from the supershell's inner edge, $R_{b}$, \citep{Draine2011,MartinezGonzalezetal2014}

\begin{eqnarray}
\label{eq:equilibrium}      \hspace{-0.4cm}
\der{}{r}\left(\frac{\mu_{ion}}{\mu_{a}} n k_{B} T\right) &=& n\sigma_{d} \frac{\left[L_n e^{-\tau}+L_i \phi\right]}{4\pi r^2 c} + 
 \frac{n^2 \beta_2L_i}{Q_0c} ~,\\
\label{eq:phi}      \hspace{-0.4cm}
\frac{Q_{0}}{4\pi r^2}\der{\phi}{r} &=&  - \beta_2 n^2  - \frac{n\sigma_{d} Q_{0}\phi}{4\pi r^2}  ~,\\     
\label{eq:tau}      \hspace{-0.4cm}
\der{\tau}{r} &=& n\sigma_{d} ~.
\end{eqnarray}

In the above equations $L_i$, $L_n$ and $Q_0$ are the cluster's ionising and non-ionising luminosity, and the number of ionising photons that emerge from the cluster per unit time; $n$ and $T$ are the supershell's gas number density and temperature, respectively; $\sigma_{d}$ is the grain absorption cross section per particle, $\phi$ is the fraction of the ionising power that gets to a surface with radius $r$, while $\tau$ is the absorption optical depth. Finally, $c$ and $\beta_2 = 2.59 \times 10^{-13}$~cm$^3$ s$^{-1}$, are the speed of light, and the recombination coefficient to the excited states of H \citep{Osterbrock1989}, respectively. The set of equations (\ref{eq:equilibrium}-\ref{eq:tau}) allows to calculate the thickness of the supershell's ionised section.

The grain cross section per particle, $\sigma_{d}$, averaged by the Planck function, $B_{\lambda}$, at blackbody temperature 
$T_\lambda$ and wavelength $\lambda$, and grain size distribution is calculated as \citep[\eg][]{Ferraraetal2017}

\begin{figure}
\includegraphics[width=\columnwidth]{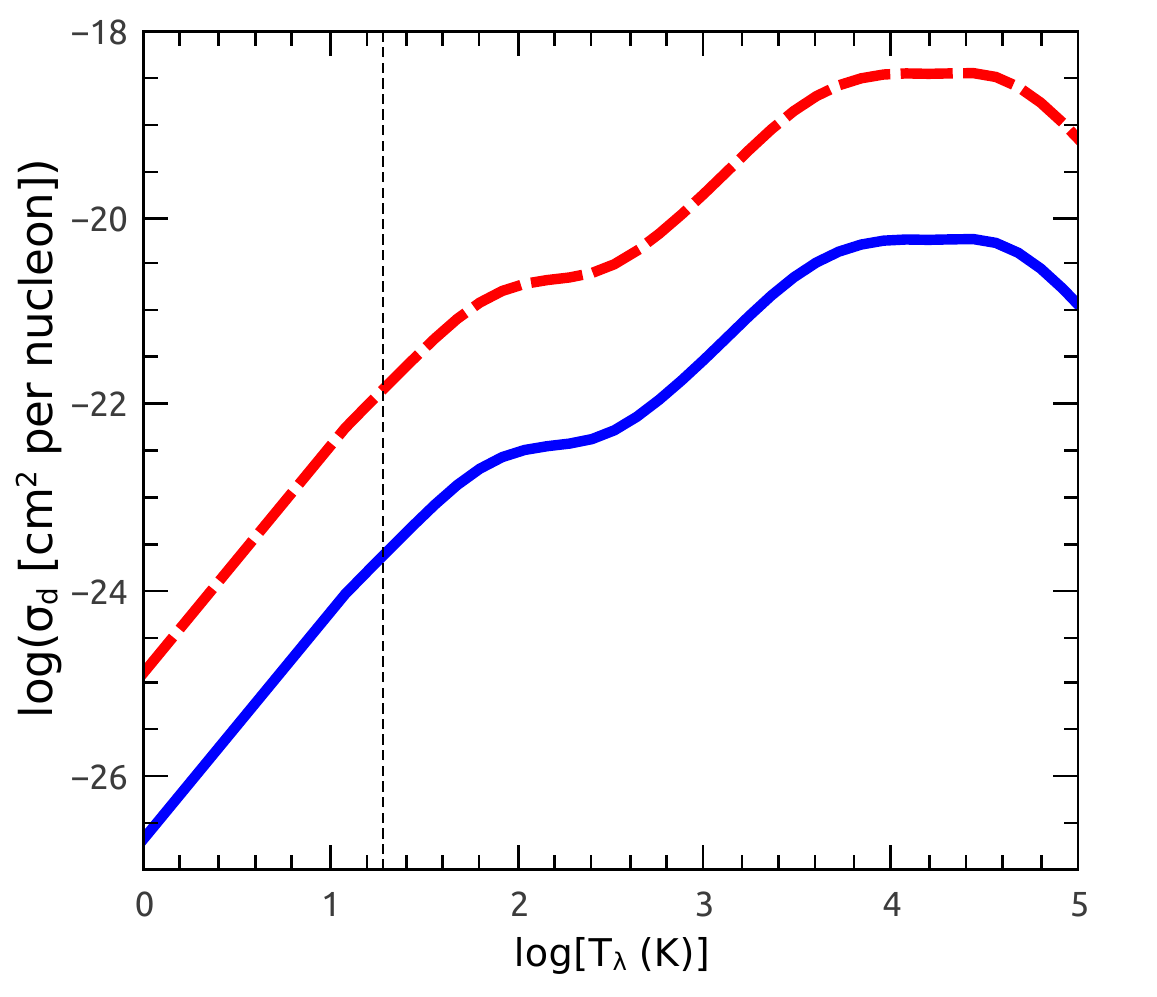}
\caption{The grain absorption cross section per particle averaged by a log-normal grain size distribution as 
a function of the blackbody temperature for carbonaceous grains and metallicity $2.5\Zsol$. The blue solid 
and red dashed lines depict $\sigma_d$ for a log-normal grain size distribution (see equation \ref{eq:gtd} 
below) with $a_0 = 0.1\mu$m, and $\delta = 0.01$ and $\delta = 0.05$, respectively. The vertical line indicates 
the value of $T_{cmb}(z=6)=19.1$ K.}
\label{fig:sigmad}
\end{figure}

\begin{eqnarray}
\label{eq:sigma}
\sigma_{d} &=& \int_{a_{min}}^{a_{max}} \int_{0}^{\infty} \frac{\pi^2 a^2 Q_{abs}(a,\lambda) B_\lambda(T_{\lambda})}{\sigma_{SB} T_{\lambda}^4} \derp{n}{a}\, \dif{a} \, \dif{\lambda} ~, 
\end{eqnarray}

where $a$ denotes the grain radius, $Q_{abs}(a,\lambda)$ is the dust absorption efficiency \citep{Draine2003,Compiegneetal2011} 
and $\sigma_{SB}$ is the Stefan-Boltzmann constant. $\sigma_{d}$ in equations (\ref{eq:equilibrium}-\ref{eq:tau}), is 
evaluated from equation \eqref{eq:sigma} at the blackbody temperature of the integrated stellar spectrum 
(see Figure \ref{fig:sigmad}).

At the supershell's inner edge, the boundary conditions are that $\tau$ is equal to zero and $\phi$ equal to unity. Given sufficient density and dust content, the ionisation front can be trapped within the supershell. This occurs when $\phi$ goes to zero. Thus a neutral skin is formed in the supershell's outer part \citep{MartinezGonzalezetal2014,Rahnetetal2017}, whose thickness is determined by the thermal pressure and the difference between the swept-up mass (see equation \eqref{eq:Mr}) and the mass of ionised gas.

\subsection{Neutral Shell}

The optical depth to CMB photons is obtained by considering that $N_{H}(r')$ is the gas column density measured from the supershell's outer edge inwards:

\begin{eqnarray}
\tau_{cmb}(r')= \sigma_{d}N_{H}(r')  ~,
\end{eqnarray}

where $\sigma_{d}$ is evaluated at $T_{\lambda}=T_{cmb}(z)$.

We have assumed a log-normal grain size distribution of the form \citep[\eg][]{HensleyandDraine2017}

\begin{eqnarray}
\label{eq:gtd}
\displaystyle \derp{n}{a} \sim a^{-1} \exp \left\{-\frac{1}{2\delta^2}\left[\displaystyle\ln\left( \frac{a}{a_{0}}\right)\right]^2\right\} ~,
\end{eqnarray}

where $a_0$ and $\delta$ set the peak grain size and width of the size distribution, and $a_{min}$ and $a_{max}$ are the lower and upper size cut-offs, respectively. The grain size distribution is normalised to the dust-to-gas mass ratio $\mathcal{D}$, which is assumed to scale with $Z_{ISM}$, as

\begin{eqnarray}
\mathcal{D} = \mathcal{\Dsol} Z_{ISM} = \frac{4\pi}{3\mu_{mol}}\rho_{gr}\int_{a_{min}}^{a_{max}}  a^3\derp{n}{a} \dif{a} , 
\end{eqnarray}

where $\mu_{mol}=2.33 m_{H}$ is the mean mass of molecular gas, $\rho_{gr}$ is the grains' bulk density (equal to $2.26$ g cm$^{-3}$ for carbonaceous grains), $Z_{ISM}$ is measured 
in solar units and $\Dsol$ is set to $1/150$ as in the solar vicinity.

\begin{figure}
\includegraphics[width=\columnwidth]{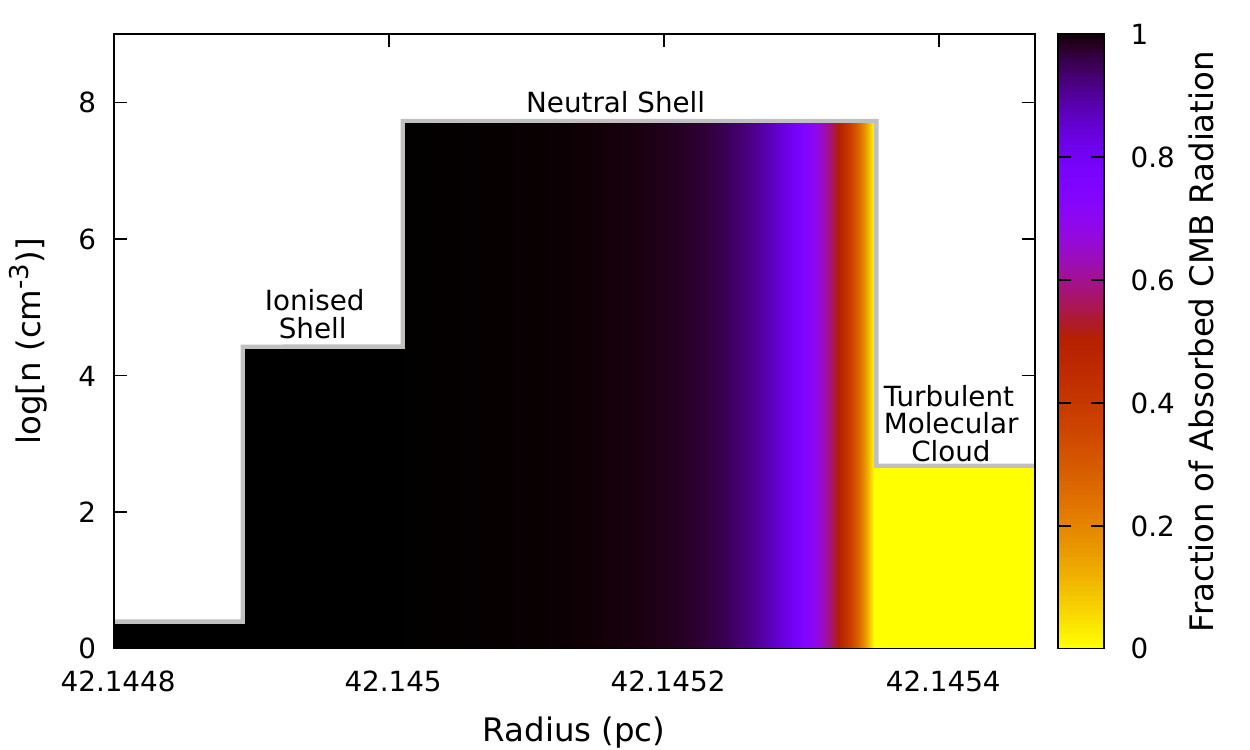}
\caption{The supershell's gas mass density structure for model A3500 at $0.5$ Myr. The supershell's ionised and neutral parts, and the cloud's unshocked part are indicated. The colour scale depicts the fraction of the CMB radiation that is trapped as it paths through the supershell.}
\label{fig:structure}
\end{figure}

\subsubsection{Hydrogen-Deuteride Molecules (HD) and Dust Cooling}

Hydrogen-deuteride molecules, efficiently formed in post-shocked gas \citep{Vasiliev1andShchekinov2005}, allow the supershell's neutral part to cool down as a result of radiation emitted from their ground state rotational transition. 
As long as the energy levels of HD molecules are populated by CMB photons, the gas cannot cool below $T_{cmb}(z)$. The 
time-scale to reach the CMB temperature via HD cooling is approximated by \citep{JohnsonandBromm2006}

\begin{eqnarray}
\label{eq:tcmb}
t_{cmb} \simeq \frac{1}{2A_{10} X_{ HD}} \left(\frac{k_{B} T_{cmb}(z)}{h\nu_{10}} \right)^2 
                                        \exp\left(\frac{h\nu_{10}}{k_{B} T_{cmb}(z)} \right) ,  
\end{eqnarray}

where $A_{10}=5\times 10^8$ s$^{-1}$ is the HD Einstein coefficient for spontaneous emission, $X_{ HD}$ is the 
fractional abundance of HD, $\nu_{10}\approx 2.7$ THz and $h$ is the Planck constant. At $z=6$, 
$t_{cmb}\approx 6.21 X_{ HD}^{-1}$ years. 

At densities $\gtrsim 10^6$ cm$^{-3}$, however, dust grains are generally the dominant coolants 
\citep{KlessenandGlover2016} and the time required to cool the neutral supershell to $T_{cmb}(z)$ can be 
significantly reduced \citep{Meeceetal2014}. The gas temperature evolution as a consequence of cooling 
via gas-grain collisions is given by

\begin{eqnarray}
\label{eq:dc}
\der{T}{t}= \frac{2\Lambda_{gas\rightarrow dust}}{5 n k_{B}}  ~,
\end{eqnarray}

where $\Lambda_{gas\rightarrow dust} \sim[1.2\times10^2$ cm$^2$ g$^{-1}$] $(\mu_{mol} n)^2 c_{s}^3$
\citep{Whitworth2016}, and $c_{s}=(k_{B} T\mu_{mol}^{-1})^{1/2}$ is the isothermal sound speed. The above equation can be integrated to obtain the time required to cool the neutral supershell down to $T_{cmb}$. The strong radiative cooling in the 
supershell leads to its rapid collapse, making it extremely thin (see Figure \ref{fig:structure}).

We calculate the supershell's inner structure using time-dependent parameters, $Q_0$, $L_i$ and $L_n$, obtained from the Starburst99 synthesis evolutionary code \citep{Leithereretal1999}. We have considered in all cases a Kroupa initial mass function sampled in the interval ($0.1-100$) $\Msol$ and Padova evolutionary models including AGB stars.  

\section{Supershell Evolution Models}
\label{sec:runs}

\begin{table*}
% \tablecaption{\label{tab:1} \sc Dusty starburst-driven supershell models}
\begin{center}
\begin{tabular}{c c c c c c c c c c c c}
\hline
\hline
{Model} & {$M_{SC}$}&{$L_{SC}$}&{$V_{\eta,\infty}$}&{$R_{c}$} & {$R_{cut}$} & {$n_{c}$} & {$\eta_{ml}$} & {$t_{ml}$} & {$M_{MC}$} & {$\epsilon$} & {$\omega$}\\
        & {\tiny $\Msol$}&{\tiny 10$^{41}$ erg s$^{-1}$} & {\tiny km s$^{-1}$} & {\tiny pc} & {\tiny pc} & {\tiny cm$^{-3}$}& {\tiny$-$} & {\tiny Myr}&{\tiny $\Msol$}  &  {\tiny $-$} &  {\tiny $-$}\\
\hline
A3500   &  $2\times10^{6}$ & $3.12$ & $3500$ & $5.0$ & $100$  & $ 3.0\times10^4$ & $0.0$ & $\infty$& $5.3\times10^7$  & $3.7\%$ & $2.0$\\
A1750   &  $2\times10^{6}$ & $3.12$ & $1750$ & $5.0$ & $100$  & $ 3.0\times10^4$ & $3.0$ &  $3.8$  & $5.3\times10^7$  & $3.7\%$ & $2.0$\\
B3500   &  $2\times10^{6}$ & $3.12$ & $3500$ & $5.0$ & $48$   & $ 3.0\times10^4$ & $0.0$ & $\infty$& $5.3\times10^7$  & $3.7\%$ & $1.5$\\
B1750   &  $2\times10^{6}$ & $3.12$ & $1750$ & $5.0$ & $48$   & $ 3.0\times10^4$ & $3.0$ &  $3.8$  & $5.3\times10^7$  & $3.7\%$ & $1.5$\\
C3500   &  $2\times10^{6}$ & $3.12$ & $3500$ & $5.0$ & $100$  & $ 3.0\times10^4$ & $0.0$ & $\infty$& $2.0\times10^6$  & $10\%$  & $2.5$\\
C1750   &  $2\times10^{6}$ & $3.12$ & $1750$ & $5.0$ & $100$  & $ 3.0\times10^4$ & $3.0$ &  $3.8$  & $2.0\times10^6$  & $10\%$  & $2.5$\\
D3500   &  $1\times10^{6}$ & $1.56$ & $3500$ & $3.0$ & $100$  & $ 5.0\times10^4$ & $0.0$ & $\infty$& $3.2\times10^6$  & $3.1\%$ & $2.0$\\
D1750   &  $1\times10^{6}$ & $1.56$ & $1750$ & $3.0$ & $100$  & $ 5.0\times10^4$ & $3.0$ &  $2.7$  & $3.2\times10^6$  & $3.1\%$ & $2.0$\\
\hline
\hline
\end{tabular}
\end{center}
\caption{The Table presents the main parameters of our six models: the star cluster mass, 
the mechanical luminosity, the adiabatic terminal speed, the cloud's core and cut-off radii, the cloud's 
gas number density at the core radius, the mass-loading factor, the mass-loading time-scale, the molecular cloud's gas mass, the star formation efficiency, and the cloud's gas density distribution power-law index.}
\label{tab:1}
\end{table*}

Our reference case A3500 (see Table \ref{tab:1}), considers a star cluster with mass $M_{SC}=2\times10^6$ $\Msol$, located within the 
central ($r<R_{c}=5$ pc), dense ($n_{c}=\rho_{c} \mu_{mol}^{-1}=3.05\times10^4$ cm$^{-3}$) zone  of a molecular cloud (see equations \ref{eq:powerlaw} and  \ref{eq:powerlaw2}), and a total mass $M_{MC}=5.3 \times 10^7\Msol$. This assumes that the molecular cloud 
follows a power-law gas density distribution with $\omega=2$ at $r\geq R_{c}$. This value is compatible with an isothermal, self-gravitating molecular cloud \citep[\eg][and references therein]{Rahneretal2019}. The molecular cloud is truncated at $R_{cut}=100$ pc and thus the star formation event occurs with a global efficiency $\epsilon \equiv M_{SC}/M_{MC}\sim3.7\%$. The metallicity in the host molecular cloud is chosen as $2.5 \Zsol$. This choice is motivated by the rapid build-up of metals in starbursting quasar-host galaxies where the  gas metallicity reaches super-solar values already at $\sim 0.1$ Gyr \citep[\eg][]{Caluraetal2014}. According to the outputs from Starburst99, the values of $L_{SC}$ and $V_{\infty}$ at the start of the cluster's evolution are $\sim3.12\times10^{41}$ erg s$^{-1}$ and $\sim3500$ km s$^{-1}$, respectively, implying a star cluster's mass loss rate of $\dot{M}_{w}\sim 8.0\times10^{-2}$ $\Msol$ yr$^{-1}$.

The fractional HD abundance is selected as $X_{ HD}=6.8\times10^{-5}$ \citep{Meeceetal2014}. This implies, from evaluation of equation \eqref{eq:tcmb}, that the neutral supershell reaches the CMB temperature in $\sim9.13\times10^4$ years via HD cooling alone. However, if one accounts for dust cooling (equation \ref{eq:dc}), the time required to reach $T_{cmb}=19.1$ K would be $\sim350$ years if the supershell's neutral section has an initial temperature $\sim10^4$ K (right after it ceases to be ionised) and a density $\sim9.4\times 10^5$ cm$^{-3}$. Moreover, dust-induced cooling is largely capable of counteracting the heating induced by H$_{2}$ formation at densities 
$\geq 10^6$ cm$^{-3}$ and metallicities $\geq 10^{-2}\Zsol$ \citep{Meeceetal2014}.

Regarding the grain size distribution defined in equation \eqref{eq:gtd}, we have fixed $\delta=0.05$, $a_0=0.1$ $\mu$m, $a_{min}=0.001$ $\mu$m and $a_{max}=0.5$ $\mu$m. For simplicity we consider only carbonaceous grains. We have imposed that supershells effectively self-shield from the CMB only when $[1-\exp(-\tau_{cmb})] \gtrsim 99.0\%$.  

For the sake of comparison, seven additional cases (A1750, B3500, B1750, C3500, C1750, D3500 and D1750) were also considered. The chosen parameters in all of them fall within the observational constraints derived for the case of a molecular cloud ($\sim 9\times 10^7\Msol$) at $z\sim6$, hosting a compact proto-globular cluster candidate with a stellar mass of a few $\sim10^6$ $\Msol$\citep{Vanzellaetal2019,Calura2021}.

Model A1750 ($\epsilon \sim3.7\%$) differs from the reference case A3500 as it includes mass-loading that results in a terminal speed $V_{\eta}=1750$ km s$^{-1}$. Models B3500 and B1750 ($\epsilon \sim3.7\%$) assume a flatter gas density distribution ($\omega=1.5$). This choice is supported by the results of \citet{Leeetal2015} and \citet{Raskuttietal2017}. Models C3500 and C1750
($\epsilon \sim10.0\%$) consider a steeper gas density distribution ($\omega=2.5$). Finally, models D3500 and D1750
($\epsilon \sim3.1\%$) explore the case of a central star cluster with $10^6\Msol$. All models denoted with 3500 have $V_{\eta}=3500$ km s$^{-1}$. Models with with 1750 correspond to mass-loaded outflows with $V_{\eta}=1750$ km s$^{-1}$, 
$\eta_{ml}=3.0$, and $t_{ml}=3.8$ Myr for models A1750, B1750 and C1750, and $t_{ml}=2.7$ Myr for model D1750.
The central zone gas mass is $M_{MC}(r<R_{c})\sim9.2\times10^5$ $\Msol$ in all but models D3500 and D1750. In those models, $M_{MC}(r<R_{c})$ is $\sim3.3\times10^5$ $\Msol$. 

In models A3500, A1750, C3500, C1750, D3500 and D1750, we have truncated the clouds at $R_{cut}=100$ pc. For models B3500 and B1750
the clouds are truncated at $R_{cut}=48$ pc so they have the same mass than in the reference case. In all models we have selected a redshift $z=6$. Our calculations are stopped once the supershells reach the edge of their host molecular cloud. The input parameters are summarised in Table \ref{tab:1}. In all the presented cases, the outward-directed force resulting from
the shock-heated gas thermal pressure overcomes the inward-directed gravitational force acting on the supershell during the whole supershell evolution: 

\begin{eqnarray}
\label{eq:Fpot}
 P_{b}  > \frac{1}{4\pi r^2}\left( \frac{GM_{sh}(r)M_0}{r^2}+\frac{1}{2}\frac{GM_{sh}^2}{r^2} \right) ~,
\end{eqnarray}

where $M_{0}=M_{MC}(r<R_{c})+M_{SC}$. This warrants that the star cluster’s mechanical energy is sufficient to overcome the binding 
energy of the star cluster \citep{Baumgardtetal2008} and form an expanding supershell.

In our reference model A3500, the supershell self-shields from the star cluster ionising and non-ionising radiation as soon as it starts to grow (\ie both ($1-\phi$) and [$1-\exp(-\tau)$] go rapidly to unity).

Figure \ref{fig:structure} presents the density structure of the starburst-driven supershell at $0.5$ Myr of evolutionary time in the reference model A3500. In this case, the fraction of CMB radiation that is absorbed within the supershell's neutral part is remarkably high, with $[1-\exp(-\tau_{cmb})]$ reaching $\gtrsim 99.9\%$, during $\sim 0.65$ Myr, and $\gtrsim 99.0\%$ during 
$\sim1.0$ Myr. In all models with mass-loading (decreased terminal speed), the supershell self-shields from the CMB radiation
for a longer time. Interestingly, if we consider a flatter gas density distribution, as in cases B3500 and B1750, the self-shielded gas mass increments monotonically, and the supershells in those cases maintain $[1-\exp(-\tau_{cmb})]\gtrsim 99.9\%$ during the whole time 
they evolve within their host molecular cloud (see Section \ref{sec:growth}).

Models C3500, C1750, D3500 and D1750 show similar trends to the ones observed in models A3500 and A1750, but the time in which $[1-\exp(-\tau_{cmb})]\gtrsim 99\%$ is reduced to $\sim 0.36$ Myr, $\sim 0.4$ Myr, $\sim 0.6$ Myr, and $\sim 0.75$ Myr, respectively. 

\begin{figure}
\includegraphics[width=\columnwidth]{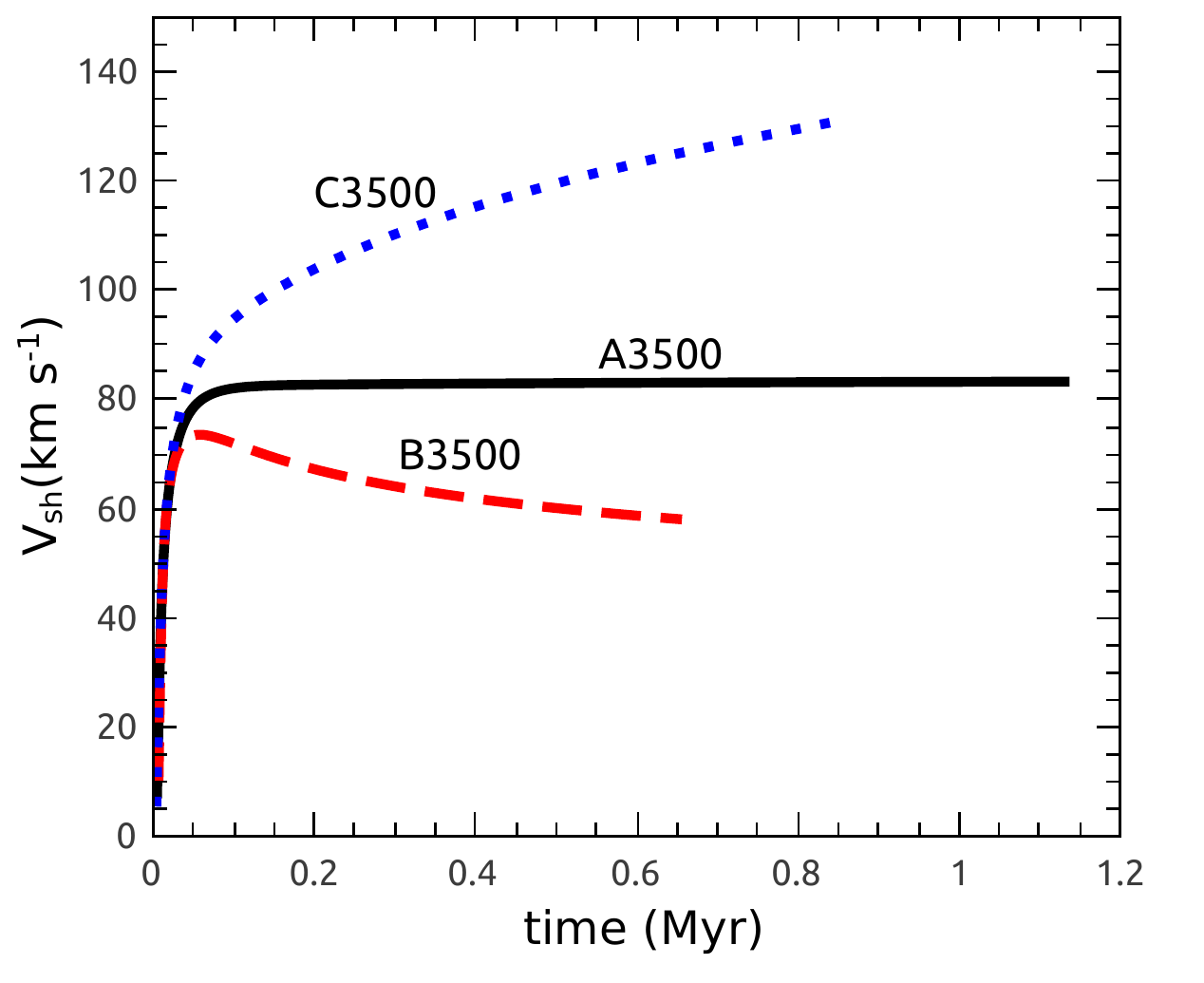}
\caption{The supershell expansion velocity in different molecular gas density distributions for models A3500 (black solid line), B3500 (red dashed line), and C3500 (blue dotted line). Note that the calculations are stopped when the supershells reach the edge of their host
molecular clouds.}
\label{fig:shellexp}
\end{figure}

The supershell expansion velocity in different ambient gas density distributions is shown in Figure \ref{fig:shellexp}, 
where solid, dashed and dotted lines correspond to models A3500, B3500 and C3500, respectively. The supershells in all our models do not stall within their host molecular clouds. However, in clouds with a power-law index $\omega > 2$, the supershells accelerate and, inevitably, become Rayleigh-Taylor unstable \citep{Krauseetal2012}. Note also that in simulations with larger mass-loading rates 
(terminal speeds $V_{\eta} < 1500$ km s$^{-1}$), catastrophic gas cooling sets in the shock-heated gas zone that prevents the formation of supershells.

We have restricted our models to the spherically-symmetric case. However, if the host molecular 
cloud is clumpy, the starburst-driven superbubble would expand preferentially through channels in between 
clumps \citep{Lucasetal2020}. We will explore that scenario in a forthcoming communication by means of three-dimensional hydrodynamical simulations.

\section{The Growth of Dust Grains}
\label{sec:growth}

As a consequence of the supershell's self-shielding against the incoming CMB/starlight radiation, the restrictions for dust grain growth at high-redshift \citep{Ferraraetal2016} are evaded. Indeed in this case:

\begin{itemize}
 \item Dust grains in supershells can cool below $T_{cmb}$. For instance, a dust grain would take just a few days to cool down to $15$ K or less (see Appendix \ref{app:1}).
 
 \item The grains will not photo-desorb their accreted species, nor will they become Coulomb repulsive.
\end{itemize}

The grain growth time-scale is given by \citep[][]{Spitzer1978}

\begin{eqnarray}
\label{eq:tgrowth}
 t_{growth} =\left[\frac{3 S \mathcal{D}\mu_{mol} n}{4\rho_{gr}a}\left(\frac{8 k_{B} T}{\pi m_{s}}\right)^{1/2}  \right]^{-1} ,
\end{eqnarray}

where $S=[0,1]$ is the sticking coefficient, and $m_{s}$ is the mass of the accreting species. The redshift-dependent CMB photon number density is given by \citep[\eg][]{Grupen2005}

\begin{eqnarray}
\label{eq:ngamma}
n_{\gamma} = 16\pi \zeta(3) \left( \frac{k T_{CMB}(z)}{ch} \right)^3  ~,
\end{eqnarray}

where $\zeta$ is the Riemann zeta function. The CMB photon number density scales with redshift as $\approx 411(1+z)^3$ cm$^{-3}$, so its value (attenuated  by dust absorptions) at $z=6$ is $\approx 1.41\times 10^5[1-\exp(-\tau_{cmb}(r'))]$ cm$^{-3}$. The dust absorption rate per unit volume for the case of CMB photons, is 

\begin{eqnarray}
 \kappa_{cmb}=n \sigma_{d} F_{\gamma}[1-\exp(-\tau_{cmb}(r'))] \, ~,
\end{eqnarray}

where $F_{\gamma}=n_{\gamma} \times c/\sqrt{3}$ is the mean CMB photon flux \citep{Heacox2017,Kimetal2019}.

\begin{figure}
\includegraphics[width=\columnwidth]{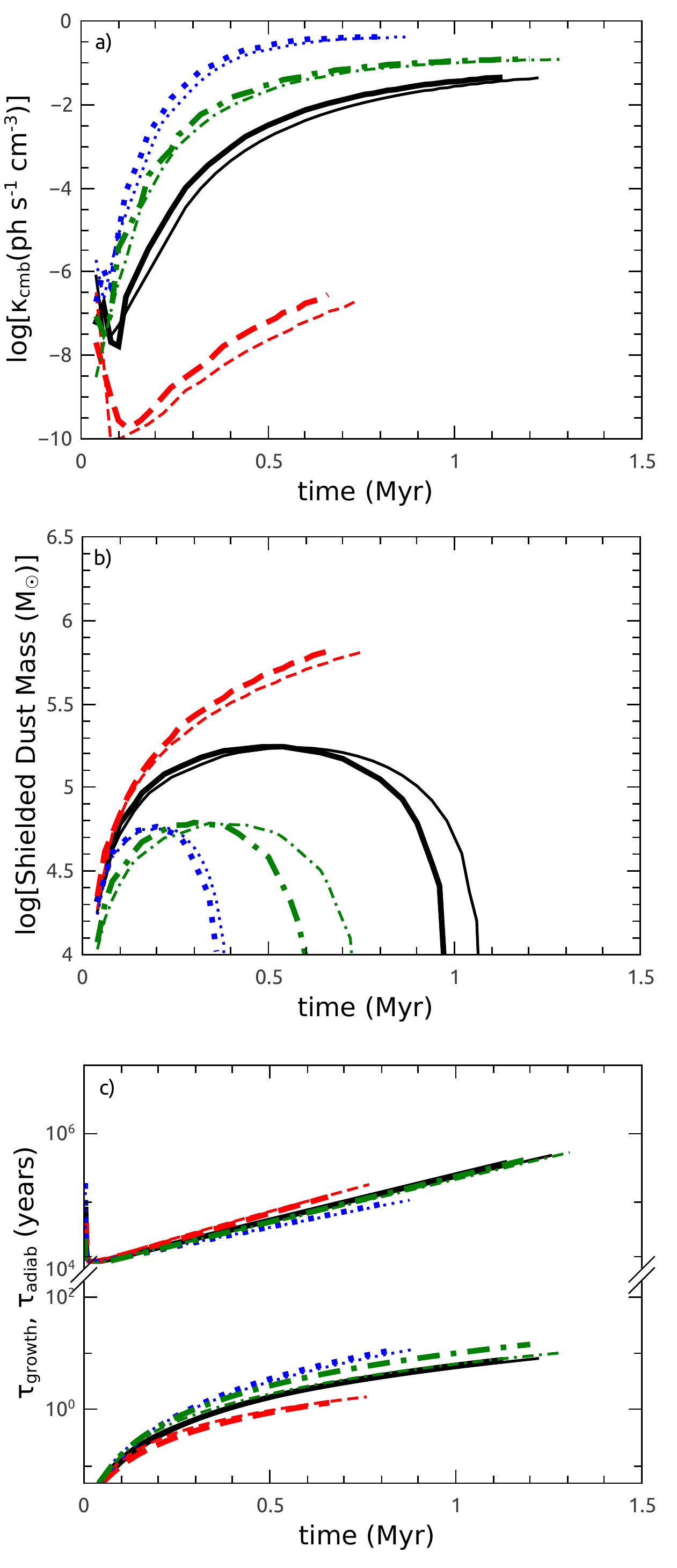}
\caption{The starburst-driven supershell properties. Panel a presents he dust absorption rate per unit volume. The amount of dust in the self-shielded region as a function of time is presented in panel b. Lines below and above the axis break on panel c display the adiabatic cooling and the dust grain growth (for $S=1$ and $a=0.01$ $\mu$m) time-scales, respectively. In each panel, we present the results for the A3500 (thick black solid lines), A1750 (thin black solid lines), B3500 (thick red dashed lines), B1750 (thin red dashed lines), C3500 (thick blue dotted lines), C1750 (thin blue dotted lines), D3500 (thick dash-dotted lines) and D1750 (thin dash-dotted lines) cases, respectively.}
\label{fig:A-C}
\end{figure}

Figure \ref{fig:A-C}, panel a, shows $\kappa_{cmb}$ calculated at the inner edge of the neutral supershell. In all cases $\kappa_{cmb}$ is tiny during much of the supershell's evolution. This implies that the CMB radiation and the gas and dust are thermally-decoupled. Panel b in the same Figure shows that, initially, the self-shielded gas mass increases as it collects more mass. However, 
depending on the cloud's radial gas density distribution, the fast supershell expansion eventually leads to a decreased supershell's density. This provokes a decline in the amount of self-shielded gas mass in the case of a steep ambient gas density distribution
(in cases A3500 and A1750 this occurs after $\sim0.5$ Myr). This decline does not occur for our models with a flatter
($\rho \sim r^{-1.5}$) gas density distribution.

The grain growth time-scale within the neutral supershell is so fast in our models (see panel c in Figure \ref{fig:A-C}) and the dust-induced cooling is so efficient, that one can expect that all the available refractory elements will inevitably be accreted onto dust grains. For instance, for $a=0.01$ $\mu$m, $S=1$, and $T=T_{cmb}(z=6)$, $t_{growth}$ is  $\lesssim 100$ years during the first $\sim2$ Myr, while if $S$ acquires a modest value, \eg 0.1, $t_{growth}$ would be $\lesssim 1000$ years during that period. Panel c also shows that adiabatic cooling (lines above the axis break, see equation \ref{eq:tadia} in Appendix \ref{app:1}) is at play, but with a reduced effect due to its significantly longer characteristic time-scale. Hence, grains can grow very rapidly in the extremely dense, heavily-enriched and cold neutral part of the dusty supershell.

\begin{figure}
\includegraphics[width=\columnwidth]{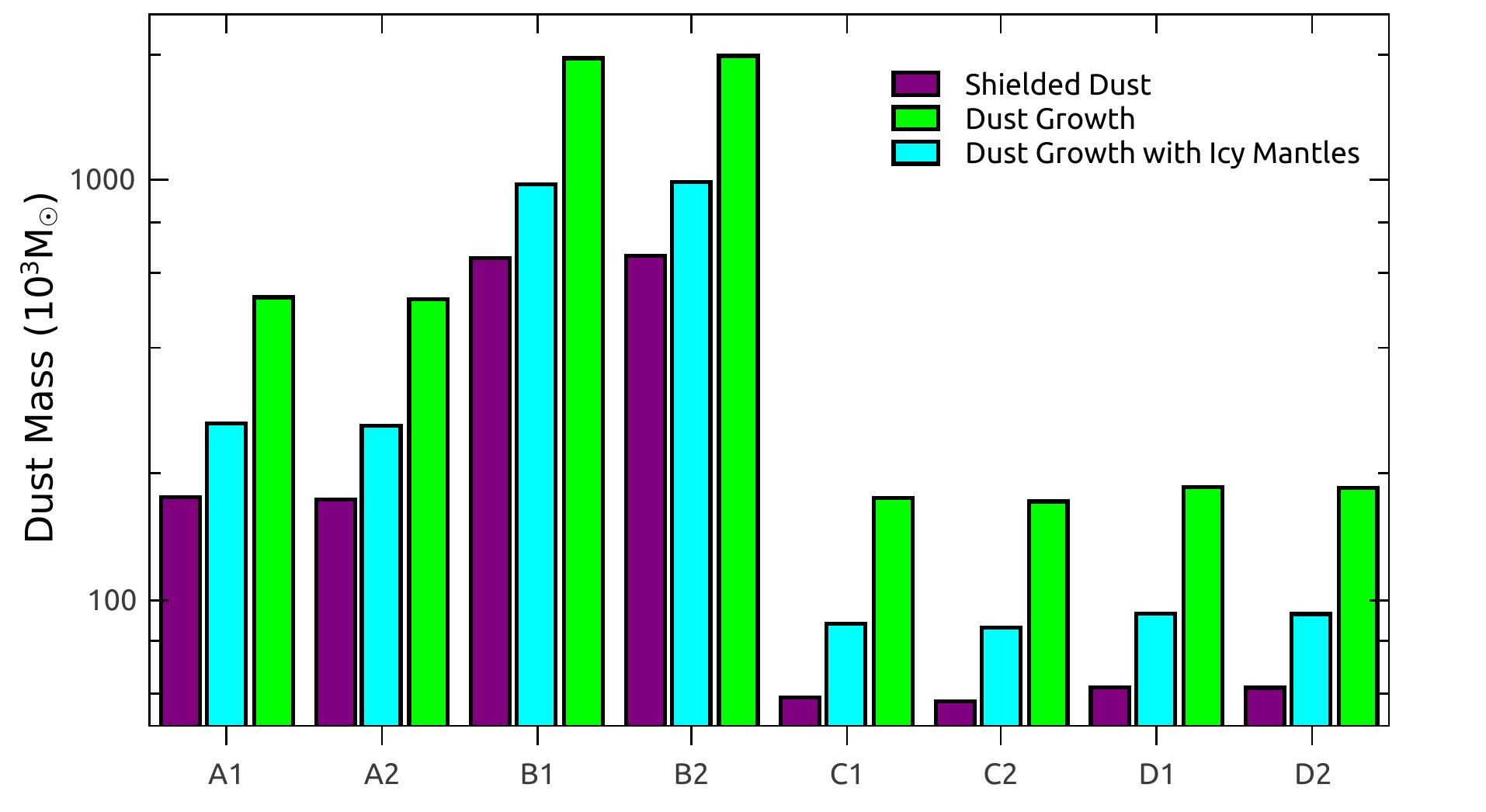}
\caption{Bar graph of the growth of dust mass in the self-shielded region for each studied case. The bars
in purple, cyan and green indicate the pre-existent swept-up dust mass in the self-shielded region (in units
$10^3\Msol$), the growth of the dust mass due to accretion of refractory elements, and the estimated growth of dust mass due to icy mantles, respectively.}
\label{fig:dustmass}
\end{figure}

\begin{figure}
\includegraphics[width=\columnwidth]{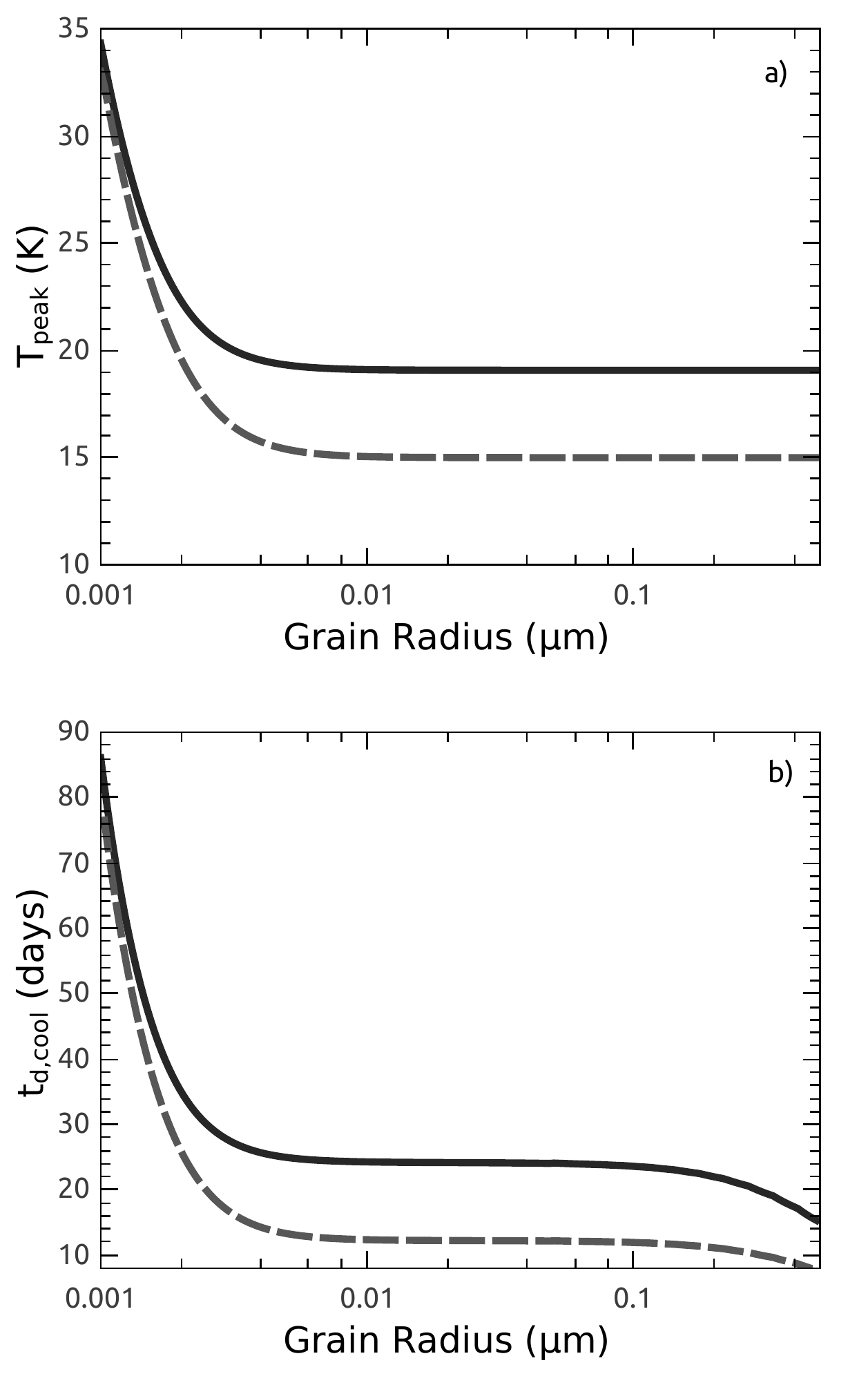}
\caption{$T_{peak}$ as a function of grain radius for the case of grains colliding with an HD-emitted 
photon (panel a). The time required to cool from $T_{peak}$ to $T_{d}=10$ K as a function of grain radius (panel b). In both 
panels the solid lines assume the initial grain temperature to be $19.1$ K, while the dashed lines depict the 
case for an initial grain temperature equal to $15$ K.}
\label{fig:Tpeak}
\end{figure}

For an assumed maximum dust-to-gas mass ratio $\sim1/100$ \citep[][]{ValentiniandBrighenti2015}
at solar metallicity, scaling with $Z_{ISM}$, the dust mass will have a net increase of $\sim8.8\times10^4$ $\Msol$ for model A3500. 
and $\sim8.6\times10^4$ $\Msol$ for model A1750. In models B3500 and B1750, the dust mass increase is $\sim3.26\times 10^5$ $\Msol$ and $\sim3.31\times 10^5$ $\Msol$, respectively. The accelerating supershells in models C3500 and C1750 rupture as they become Rayleigh-Taylor unstable, and consequently, it is uncertain whether dust grain growth could occur. Nevertheless, we report the results of our calculations for these models in Figure \ref{fig:dustmass}. Finally, the dust mass increments are $\sim3.1\times 10^4$ $\Msol$ for both models D3500 and D1750. The above quantities disregard the dust mass in the form of ice mantles, which can further boost the grains' mass. The bars in Figure \ref{fig:dustmass} represent the amount of dust that was originally swept-up (purple), the amount of dust that is obtained after all refractory elements are locked-up onto the grains (cyan), and the grain mass in the form of ice mantles, which is assumed to increase the total dust mass by a factor of two \citep{Krugel2003}. As a result, the supershells' self-shielded sections 
will have, within $\sim1$ Myr, three times more dust mass than the original self-shielded dust mass.

As expected, the most successful cases are those in which the stellar masses are larger. If one assumes a cluster mass function 
${\rm{dN}} / {\rm{d}} M_{\rm{*}} \propto M_{\rm{*}}^{\beta}$, with $\beta \approx -2$ \citep[\eg][]{Cooketal2019}, in a galaxy with lower and upper cluster masses $10^3 \Msol$ and $10^7\Msol$, respectively, then the number of clusters with masses $\gtrsim 10^6\Msol$ scales as $\sim10 (M_{*}/10^{8}\Msol)$. In the above $M_{*}$ stands for the total stellar mass. If the typical net increment in dust mass within CMB-dark supershells in clusters with $\gtrsim 10^6\Msol$ is $\sim10^5\Msol$ (and all of them reside within dense molecular clouds), then the proposed mechanism can account for the build-up of $\sim10^6\Msol (M_{*}/10^{8}\Msol)$ of dust.

This rough estimate takes into account that the fraction of stars that form in bound clusters grows with 
redshift, and at $z\sim6$ they likely made the larger proportion \citep[e. g.][and references therein]{Vanzellaetal2019}.

We have seen so far that dust grain growth within starburst-driven supershells can be both very rapid and very efficient in time-scales of order $\lesssim 1$ Myr, before any supernova goes off in the central starburst. However, between $\sim3-40$ Myr, the massive stars in the clusters will explode as core-collapse supernovae. In that scenario, supernova blast waves will ram through the hot gas cavity and collide with the encompassing supershell. Such situation was studied by \citet{MartinezGonzalezetal2018,MartinezGonzalezetal2019} by means of three-dimensional hydrodynamical simulations. Upon the collision, the supernova remnants become strongly radiative and do not experience the Sedov-Taylor phase \citep[see Figure 5 in][]{MartinezGonzalezetal2019}. Moreover, blast waves are mostly 
reflected and soon catch up with the reverse shock as they only penetrate a very thin layer into the 
supershell \citep{TenorioTagleetal1990}. Thus, the dust grains locked-up in the supershell will remain largely 
unaffected.
 
Supershells may eventually become pressure-confined, perhaps outside their host molecular clouds, and the time required for them to stall is shorter than the life-time of massive stars and their continuous winds (several tens of Myr). This implies that the central cluster will continue to produce a high velocity outflow ($\gtrsim 1000$ km s$^{-1}$). The dense ($\gtrsim 10^6$ cm$^{-3}$) stalling supershell rests then above a lower density ($\lesssim 10^{-1}$ cm$^{-3}$), hot ($\gtrsim 10^7$ K) gas. In that case the standing supershell feels an inward gravity and becomes Rayleigh-Taylor unstable. Consequently, the blast waves will collide with supershell fragments and the hot ejecta will pass through channels in between them, establishing pressure equilibrium. The blast waves will then attempt to ram through the fragments, but one can expect that they will only be weakly transmitted. Thus the large majority of the supershell's dust grains will survive to eventually mix with the unshocked ISM.

This situation differs from that described by \citet{Weaveretal1977} (our models with $\omega \leq 2.0$), where decelerating supershells are stable because the hot gas rests on top of the dense expanding supershell, and then a co-moving parcel of fluid feels an effective {\it outward} gravity.

\section{Concluding Remarks}
\label{sec:conclusions}

Based on semi-analytic models, we have analysed the large-scale evolution of starburst-driven supershells and the micro-physics of the dust grain growth process within supershells at high-redshift ($z\sim6$).

The main model predictions are summarised as follows: 

\begin{itemize}
 
 \item Supershells can self-shield from the stellar radiation field and from the Cosmic Microwave Background. The latter requires them to be sufficiently metal-rich ($Z\geq\Zsol$), dense ($n\geq 10^6$ cm$^{-3}$) and dusty ($\mathcal{D}\sim 1/150 \times Z$), and that the gas density distribution in the host molecular cloud is not too steep ($\omega \leq 2$).

 \item As a result of dust-induced and HD cooling (and less importantly adiabatic expansion), the supershell's dust grains can cool below the temperature set by the CMB ($T_{cmb}(z=6)=19.1$ K), \eg to a temperature $\sim15$ K or less within a few days.

\item The grain growth time-scale in the self-shielded section of supershells is sufficiently short ($\lesssim 100$ years) to allow all refractory elements within the supershell's self-shielded section to be accreted onto dust grains. 

\item These grains do not photo-adsorb the accreted species, nor become Coulomb repulsive, and thus can survive being 
self-shielded from the central cluster starlight.
      
\item The total amount of dust produced by this mechanism may reach $\sim10^6\Msol (M_{*}/10^{8}\Msol)$.
\end{itemize}

This mechanism takes places before the first supernova explosion in the central star cluster ($\sim3$ Myr). Nevertheless, supernovae do not pose a significant threat to the survival of the dust grains locked-up within the encompassing supershell because the latter mostly reflects the colliding blast waves, leaving its dust content largely unaffected \citep{MartinezGonzalezetal2019}. We have thus shown that dust grain growth at high redshift may take place rapidly and efficiently.

\section{Acknowledgements}
This study was supported by CONACYT-M\'exico research grant A1-S-28458. S.M.G. also acknowledges support by CONACYT 
through C\'atedra n.482. The authors thankfully acknowledge the computer resources, technical expertise and support 
provided by the Laboratorio Nacional de Superc\'omputo del Sureste de M\'exico, CONACYT member of the network of national 
laboratories, and by the Laboratorio Nacional de C\'omputo de Alto Desempe\~{n}o (LANCAD), project
13-2021. The authors thank the anonymous Referee for a careful reading and helpful suggestions which 
greatly improved the paper. 

\section{Data Availability}

The data underlying this article will be shared on reasonable request to the corresponding author.

\bibliographystyle{mnras}
\bibliography{Infrared}

\appendix

\section{Turbulent Molecular Cloud}
\label{app:turb}
The pressure gradient in the static initial cloud is determined by the equation \citep[\eg][]{Caluraetal2015}

\begin{eqnarray}
\label{eq:hydrostatic}
 \der{P_t}{r}=\frac{-GM(r)\rho}{r^2} ~,
\end{eqnarray}

where $G$ is the gravitational constant. In the case of a power-law density distribution 
(see equation \ref{eq:powerlaw}), the mass enclosed within a radius $r$, $M_{MC}(r)$, is 

\begin{eqnarray}
\label{eq:Mr}
 M_{MC}(r) = M_{0} + \frac{4\pi\rho_{c}R_{c}^3}{(3-\omega)}\left[\left(\frac{r}{R_{c}}\right)^{3-\omega}-1 \right]  ~,
\end{eqnarray}

where $M_{0}$ is the total (the star cluster and the residual gas) mass within the central zone with 
radius $r<R_{c}$. The second term in equation \eqref{eq:Mr} gives the mass contained in the swept-up 
supershell, $M_{sh}$.

In the case of a power-law density distribution, equation \eqref{eq:hydrostatic} is easily integrated:

\begin{eqnarray}
 P_t&=& \displaystyle \frac{G M_{0}\rho_{c}}{(\omega+1)R_{c}}\left(\frac{r}{R_{c}}\right)^{-(\omega+1)}+
 \frac{2\pi G \rho_{c}^2 R_{c}^2}{(3-\omega)(\omega-1)} \nonumber \\
 &\times& \left[\left(\frac{r}{R_{c}}\right)^{2(1-\omega)}-
 \frac{2(\omega-1)}{\omega+1}\left(\frac{r}{R_{c}}\right)^{-(\omega+1)} \right] ~,
\end{eqnarray}

if $\omega>1$ and $P_t(\infty)=0$.

\section{Heating and Cooling in the Self-Shielded Neutral Shell}
\label{app:1}

Dust grains residing within the supershell's self-shielded, neutral part do not experience
stochastic temperature fluctuations due to collisions with CMB photons, however, they still have temperature fluctuations 
as they collide with neutral species and/or absorb HD-emitted ($h\nu_{10}\sim1.1\times10^{-2}$ eV) and infrared re-emitted photons. After each collision/absorption event, a dust grain with initial temperature $T_{0}$ will be heated to a peak temperature, $T_{peak}$, given by 

\begin{eqnarray}
E = \int_{T_{0}}^{T_{peak}} C(a,T_{d}) \mbox{ d}T_{d} .
\end{eqnarray}

where $C(a,T_{d})$ is the grain heat capacity as a function of the grain's radius, chemical composition and 
temperature, $T_{d}$. Upon reaching $T_{peak}$, the grain starts to cool down in a time-scale given by 
\citep{Dwek1986,MartinezGonzalezetal2016,MartinezGonzalezetal2017}

\begin{eqnarray}
t_{d,cool} = \int_{T_{d}}^{T_{peak}}
              \frac{C(a,T_{d})\mbox{ d} T_{d}}{|4\pi a^2 \sigma_{SB} \langle Q_{abs} \rangle T_{d}^{4}|} ,  
\end{eqnarray}

where $\langle Q_{abs} \rangle$ is the Planck-averaged grain absorption efficiency. Figure \ref{fig:Tpeak} 
shows $T_{peak}$ and $t_{d,cool}$ as a function of grain radius for the case of grains colliding with 
an HD-emitted photon, whose energy roughly corresponds with the energy of an infrared photon with wavelength 
$\sim100$ $\mu$m. We have assumed values of $T_0$ equal to $15$ K and $19.1$ K. A significant temperature 
increase will be expected only for grains $a\leq 0.003$ $\mu$m, while $t_{d,cool}$ between $T_{peak}$ and 
$T_{d}=10$ K takes only some tens of days.

On the other hand, the supershell cools as it expands and the characteristic time-scale for adiabatic cooling is \citep{Badjinetal2016}

\begin{eqnarray}
\label{eq:tadia}
t_{adiab} = \frac{(\gamma+1)}{6(\gamma-1)}\frac{R_{S}}{V_{S}} ,  
\end{eqnarray}

where $R_S$ and $V_S$ are the forward shock radius and velocity.
  
% Don't change these lines
\bsp	% typesetting comment
\label{lastpage}
\end{document}